\documentclass[fleqn,usenatbib]{mnras}

\usepackage{newtxtext,newtxmath}
\usepackage[T1]{fontenc}

\DeclareRobustCommand{\VAN}[3]{#2}
\let\VANthebibliography\thebibliography
\def\thebibliography{\DeclareRobustCommand{\VAN}[3]{##3}\VANthebibliography}

\usepackage{graphicx}	% Including figure files
\usepackage{amsmath}	% Advanced maths commands
\usepackage{xcolor}

\newcommand{\karmma}{\texttt{KaRMMa}}
\newcommand{\karmmatwo}{\texttt{KaRMMa~2.0}}
\newcommand{\karmmaone}{\texttt{KaRMMa~1.0}}
\newcommand{\Nside}{N_{\rm side}}
\newcommand{\lmax}{\ell_\mathrm{max}}
\newcommand{\knl}{k_\mathrm{NL}}
\newcommand{\lnl}{\ell_\mathrm{NL}}

\title[\karmmatwo]{\karmmatwo\ - Kappa Reconstruction for Mass Mapping}

\author[Fiedorowicz et al.]{
Pier Fiedorowicz$^{1}$\thanks{E-mail: pierfied@arizona.edu},
Eduardo Rozo$^{1}$,
Supranta S. Boruah$^{2}$
\\
$^{1}$Department of Physics, University of Arizona, Tucson, AZ 85721, USA\\
$^{2}$Steward Observatory, University of Arizona, Tucson, AZ 85719, USA\\
}

\date{Accepted XXX. Received YYY; in original form ZZZ}

\pubyear{2015}

\begin{document}
\label{firstpage}
\pagerange{\pageref{firstpage}--\pageref{lastpage}}
\maketitle

\begin{abstract}
We present \karmmatwo, an updated version of the mass map reconstruction code introduced in \citet{karmma}. \karmma{} is a \it full-sky \rm Bayesian algorithm for reconstructing weak lensing mass maps from shear data. It forward-models the convergence field as a realization of a lognormal field.  The corresponding shear map is calculated using the standard Kaiser--Squires transformation, and compared to observations at the field level. The posterior distribution of maps given the shear data is sampled using Hamiltonian Monte Carlo chains. Our work improves on the original algorithm by making it numerically efficient, enabling full-sky reconstructions at $\approx 7$ arcmin resolution with modest computational resources. These gains are made with no loss in accuracy or precision relative to \karmmaone.  We compare the \karmmatwo\ posteriors against simulations across a variety of summary statistics (one-point function, two-point functions, and peak/void counts) to demonstrate our updated algorithm provides an accurate reconstruction of the convergence field at mildly non-linear scales. Unsurprisingly, the lognormal model fails as we approach non-linear scales ($\ell \gtrsim 200$), which in turn biases the map posteriors.  These biases are at the 2\% level in the recovered power spectrum, and at the 5\% to 15\% level for other statistics, depending on the resolution.
\end{abstract}


%%%%%%%%%%%%%%%%%%%%%%%%%%%%%%%%%%%%%%%%%%%%%%%%%%

%%%%%%%%%%%%%%%%% BODY OF PAPER %%%%%%%%%%%%%%%%%%

\section{Introduction}

Weak gravitational lensing provides one of the most powerful methods for studying the mass distribution of the Universe and its large-scale structure. Because weak lensing is sensitive to all forms of matter, it allows us to measure the total matter density field of the Universe, including dark matter. As such, all modern photometric redshift surveys include weak lensing measurements as one of their key cosmological probes \citep{hikage_etal19, Heymans_2021, y3_cosmology}. These current generation surveys measure the shapes of hundreds of millions of galaxies, while next-generation surveys like LSST and Euclid will observe billions of galaxies. Optimally constraining cosmology from these observations is critical, particularly as we are currently facing a $\sim2-3\sigma$ tension in cosmological constraints between observations of the early and late Universe \citep{Heymans_2021}.

Nonlinear structure growth in the late Universe causes the density field to become highly non-Gaussian. However, traditional cosmology analyses have generally relied upon constraining cosmology with two-point statistics, which fully characterize the statistical properties of field if and only if the field is Gaussian. Indeed, previous work has demonstrated that these non-Gaussianities contain significant additional cosmological information \citep{Gatti_2020}. For this reason, many different methods have been developed for extracting non-Gaussian information from weak gravitational lensing data. A brief list of these methods includes peak and void count statistics \citep{Dietrich_2010, Kratochvil_2010, Peel_2017, Shan_2017,desy3_peaks}, bispectra \citep{takada_bispectrum, fu_bispectrum, jung2021integrated}, higher order moments \citep{Peel_2017, Gatti_2020}, Minkowski functionals \citep{Kratochvil_2012, Petri_2013, Vicinanza_2019}, machine-learning \citep{Gupta_2018, Fluri_2018, Ribli_2019, Jeffrey_2020_inference}, and scattering transforms \citep{scattering_transforms}.

Many methods for extracting non-Gaussian information generally assume that the summary statistics of interest is calculated on weak lensing mass maps made from the shear data. Traditionally the convergence field is recovered through the Kaiser--Squires reconstruction \citep{KaiserSquires}. However, the quality of the Kaiser--Squires reconstruction is severely degraded by masking effects and survey noise. As a result, a number of mass map reconstruction techniques have been proposed \citep{wiener, Jeffrey_2018, starck2021weak, Jeffrey_2020, Porqueres_2021, remy2020probabilistic, alsing_mass_mapping}. However, the goal of reconstructing mass maps that are statistically indistinguishable from simulated maps has remained elusive. Even with improved mass map reconstructions, summary statistic-based approach are inherently sub-optimal: any summary statistics based on a mass map will necessarily lose some of the information contained in the map itself.

An optimal approach for cosmology from weak lensing surveys would rely on field-based inference.  That is, instead of trying to model a lossy summary statistic of the field, one would forward-model the full observed shear field. Initial estimates of the potential gain from such methods were remarkably high \citep{Porqueres_2021}, but have since been revised downwards significantly for vanilla (i.e., $\Lambda$CDM) cosmologies \citep{supranta}.  However, in an upcoming work, we will demonstrate that the factor-of-two improvement in cosmological constraints enabled by field-based inference is restored in $w$CDM cosmologies. Perhaps most importantly, forward modeling of the lensing field opens the door to modeling systematics at the map level \citep{fbi_foreground_contamination, fbi_intrinsic_alignment}, potentially enabling internal self-calibration methods for ameliorating systematics \citep{pyne_joachimi_21}.

Critical to field-based inference is a forward model of the shear field from cosmological parameters.  The first successful implementation was that of \cite{alsingetal17}, who used a Gaussian prior to model the weak lensing fields. \cite{Porqueres_2021} used a simulation-based technique to non-linearly evolve the density field from its initial conditions to the present day, and then integrated along the line of sight to produce the convergence field. Each approach has advantages and disadvantages: the \cite{alsingetal17} model is numerically simpler, but fails to account for non-linear physics in the generative model, while the \cite{Porqueres_2021} analyses is computationally challenging. Specifically, modeling the full 3D density field is computationally expensive, making the simulation approach especially challenging to extend to wide angle surveys.  In short, delivering on the promise of field-based inference requires the a fast, \it full-sky \rm forward model capable of capturing the impact of non-linear structure growth.

In our previous work we presented \karmma{}, a fully Bayesian framework for forward modeling the shear field and improving mass map reconstruction using a fast approximate model of the convergence field \citep{karmma}. Prior work had demonstrated that the weak lensing convergence field can be accurately modeled as a lognormal random field \citep{Taruya_2002, des_lognorm, flask} at mildly nonlinear scales. In addition, \cite{supranta} demonstrated that using the lognormal model for field-level cosmology analyses results in notable gains over using a Gaussian model. For these reasons, \karmma\ models the convergence field as a lognormal random field. This first iteration of \karmma{} -- \karmmaone{} -- successfully demonstrated the application of the lognormal model in a forward modeling-based approach for mass map reconstruction \cite{karmma}.

In this work, we present \karmmatwo, an updated version of \karmma. \karmmatwo{} improves on \karmmaone{} by enabling full-sky mass map reconstructions at high resolution, a problem that was previously numerically intractable.  We demonstrate that because our updated code can now operate at small, highly nonlinear scales, the failure of the log-normal model to fully capture non-linear growth now impacts mass map reconstruction at detectable levels. Nevertheless, \karmmatwo{} outperforms existing mass map reconstruction techniques, while being the only Bayesian reconstruction algorithm capable of running on wide-field surveys.

This paper is laid out as follows. In Section \ref{sec:method} we briefly review weak lensing theory and the lognormal model. We then detail our improvements over \karmmaone{}. In Section \ref{sec:sim_tests} we outline how we test \karmmatwo{} using simulations and present our results. Subsequently, we discuss how \karmma{}'s performance compares to other reconstruction techniques. Finally, we summarize this work and present our conclusions in Section \ref{sec:conclusion}.

\section{The \karmmatwo\ Algorithm}
\label{sec:method}

\subsection{Weak Lensing Formalism}

Mass along the line of sight slightly distorts the observed shapes of source galaxies, resulting in magnification and shearing of the original image. For a perfectly circular source, the image becomes elliptical with ellipticity
\begin{equation}
    \label{eq:elip}
    \epsilon = \frac{\gamma}{1 - \kappa}.
\end{equation}
The convergence $\kappa$ parametrizes the magnification of the source, and the shear is parametrized by the complex value $\gamma$. In the weak lensing limit, $\kappa \ll 1$ and therefore $\epsilon \approx \gamma$.  That is, by measuring the ellipticity of galaxies, we can recover a noisy estimate of the lensing shear field.  In practice, we cannot measure the shear of a single source, as we do not know the shape of the undistorted source image. However, by averaging over a large number of sources (whose original ellipticities we assume are unaligned), we can measure the average value of the shear field in a given region of the sky.  Finally, since convergence and shear are related, measurements of the shear enables us to infer the convergence field of the Universe.

The weak lensing shear and convergence field are both sourced by the distribution of matter along the line of sight to a source.  In the Born approximation, the relation between these fields and the mass distribution of the Universe along a line of sight $\vec\theta$ is 
\begin{equation}
    \kappa(\vec{\theta}) = \int_0^\infty g(\chi) \delta(\chi, \vec{\theta}) d\chi
\end{equation}
where $\chi$ is the comoving distance, and $\delta = \rho/\bar\rho-1$ is the matter density contrast. The lensing kernel $g(\chi)$ determines the sensitivity of the convergence to the density field as a function of distance, and is set by the source redshift distribution $p(\chi)$,
\begin{equation}
    \label{eq:lens_kernel}
    g(\chi) = \frac{3 H_0^2}{2 c^2} \Omega_m \int_\chi^\infty p(\chi') \frac{\chi (\chi' - \chi)}{\chi'} d\chi'.
\end{equation}
Note that in this equation we have assumed a flat Universe. 

The shear field is related to the convergence field through the Kaiser--Squires relationship \citep{KaiserSquires}. On the spherical sky, this relationship is most easily expressed in terms of the spherical harmonics \citep{spherical_KS},
\begin{equation}
    \gamma_{\ell m} = - \sqrt{\frac{(\ell - 1) (\ell + 2)}{\ell (\ell + 1)}} \kappa_{\ell m},
    \label{eq:KSrel}
\end{equation}
where $\gamma_{\ell m}$ and $\kappa_{\ell m}$ are the spherical harmonic components of the shear and convergence fields. Note that the shear field is a spin-2 field and requires spin-2 spherical harmonic transformations. This relationship allows one to predict the shear field from the convergence field. 
In the absence of a data mask and noise, one can reconstruct the convergence field from the shear field by simply inverting this relationship.  Indeed, this approach is often taken in practice \citep[e.g.,][]{y3_mass_map, Oguri_2017} even though the resulting mass maps can be severely biased (as demonstrated in \citealt{y3_mass_map} and \citealt{karmma}). The \karmmatwo{} algorithm avoids these difficulties by forward modeling the shear field based on a lognormal model for the lensing convergence.

\subsection{The Lognormal Model}

The lognormal model assumes that the convergence field can be written as a simple non-linear transform of a Gaussian field.  That is, the pixelized convergence field $\kappa_i$ can be expressed as 
\begin{equation}
    \label{eq:lognorm}
    \kappa_i = e^{y_i} - \lambda,
\end{equation}
where $y_i$ is a Gaussian random field and $\lambda$ is the so-called ``shift'' parameter. The correlation function of $y_i$ is related to the correlation function of the convergence field via
\begin{equation}
    \label{eq:two_pt_relation}
    \xi_{yy}(\theta) = \log \left[ \frac{\xi_{\kappa\kappa}(\theta)}{\lambda^2} + 1 \right].
\end{equation}
The mean of $y_i$ is set by the constraint that the average convergence field is zero ($\left< \kappa_i \right> = 0$).  One finds
\begin{equation}
    \langle y_i \rangle = \ln \lambda - \frac{\xi_{yy}(0)}{2}.
\end{equation}
The shift parameter, $\lambda$, and correlation function, $\xi_{\kappa \kappa}(\theta)$, are specified by the cosmological parameters. The former can be fit from simulations\footnote{The shift parameter can also be computed using a code like \textsc{CosMomentum} \citep{Friedrich_2020}.} and the latter using tools like \textsc{CAMB} \citep{Lewis_2000} or \textsc{CLASS} \citep{Blas_2011}. We use \textsc{PyCCL} \citep{ccl} which calls \textsc{CAMB} to compute the non-linear matter power spectrum using halofit \citep{halofit_orig, halofit_revised}.  When applying our code to the mocks described in Section \ref{sec:sim_tests} we then use the prescription from \cite{hsc_mocks} to correct the power spectrum for simulation artifacts arising from the ray-tracing scheme and the finite angular resolution of the mocks. After applying the provided correction, we find residual artifacts remain at the few percent level, which we remove by fitting a small multiplicative correction to the power spectrum measured.  The necessary bias is fit using a regularized polynomial regression (polynomial order 5, regularization $\alpha=0.5$). For more information on the lognormal model, we refer the reader to \cite{flask}.

\subsection{Improvements Over \karmmaone}

In \cite{karmma}, we characterized the convergence field in configuration space: the value of the field $y$ at each pixel was a model parameter, and a Gaussian prior was imposed on the resulting map.  This approach suffers from a critical deficiency: the covariance matrix of the resulting mass map is both extremely large and dense (i.e. not diagonal).  To invert this matrix, \karmmaone\ relied on Singular Value Decomposition (SVD), an $O(n^3)$ operation. Our algorithm quickly became compute and memory limited at a resolution of $\approx 14$~arcmin in a DES Y1-footprint  ($\approx 28$~arcmin for a DES Y5-footprint).

In this work, we switch from using the real space values of the pixelized field $y$ to using the corresponding spherical harmonic coefficients.  Thanks to the homogeneity and isotropy of the Universe, these harmonics are uncorrelated.  The corresponding covariance matrix is therefore diagonal, with the entries corresponding to the power-spectrum of the field: $\langle \kappa_{\ell m}\kappa_{\ell' m'} \rangle = C_\ell \delta_{\ell \ell'} \delta_{m m'}$.  In this approach, the most computationally expensive operations performed as part of the modeling pipeline are the spherical harmonic transforms, which scale as $O(n^{3/2})$.  Consequently, \karmmatwo\
easily outperforms \karmmaone\ in run time while benefiting from a drastically smaller memory footprint.  Moreover, by modeling the spherical harmonic coefficients we reconstruct the convergence map across the full spherical sky, thereby eliminating masking artifacts in our reconstruction.

In practice, our model parameters are not the $y_{lm}$ themselves, but rather the quantities
\begin{equation}
    x_{lm}=\frac{y_{lm}}{\sqrt{C_l}}.
\end{equation}
Consequently, the set of spherical harmonics $\{x_{lm}\}$ is normally distributed with a unit covariance matrix,
\begin{equation}
    P(\vec{x}) \propto \exp \left[-\frac{1}{2} \vec{x}^\top \mathbf{I} \vec{x}  \right].
\end{equation}
This parametrization helps decorrelate the cosmological parameters from the relative phases and amplitudes of $\kappa$ \citep{supranta}. While \karmmatwo\ does not yet allow for marginalization over cosmological parameters when inferring mass maps, we intend to include this feature in future releases of the code.

\subsection{Likelihood Model and Parameter Sampling}

We model the observed shear as a noisy version of the true shear field.  The latter is related to the convergence field via equation~\ref{eq:KSrel}. We assume the source density is large enough that the shape noise in each spatial pixel can modeled using a Gaussian distribution. The full posterior distribution in our model is therefore
\begin{equation}
\begin{split}
    P(\vec{x} \mid \vec{\gamma}_\mathrm{obs}) &\propto P(\vec{x}) \times P(\vec{\gamma}_\mathrm{obs} \mid \vec{x}) \\
    &\propto P(\vec{x}) \times \exp \left [ -\frac{1}{2} \sum_i \frac{
    (\gamma_{i,\mathrm{obs}} - \gamma_i(\vec{x}))^2}{\sigma_{\gamma, i}^2} \right].
\end{split}
\end{equation}
To compute $\gamma(\vec{x})$, we proceed as follows:
\begin{enumerate}
    \item For each spherical harmonic, set $y_{lm}=\sqrt{C_l}x_{lm}$.
    \item Perform a spherical harmonic transform to get $y$ in pixel-space.
    \item Perform the lognormal transformation to arrive at $\kappa$.
    \item Use the Kaiser-Squires relation to obtain $\gamma$ via spherical harmonic transforms.
\end{enumerate}

We sample the $x_{lm}$ parameters using the \textsc{Pyro} \citep{pyro} probabilistic programming language. \textsc{Pyro} takes advantage of \textsc{PyTorch} for autograd calculations and GPU acceleration, enabling efficient Hamiltonian Monte Carlo \citep{neal2012mcmc} chain sampling. To use \textsc{Pyro} with the HEALPix spherical harmonic transforms, we implemented custom \textsc{PyTorch} autograd functions\footnote{\url{https://pytorch.org/docs/stable/notes/extending.html\#extending-torch-autograd}}, which call the underlying HEALPix routines for the forward and backward (gradient) calculations. Unfortunately, we cannot take advantage of GPU acceleration because the HEALPix spherical harmonic routines do not have a GPU implementation. Should such routines become available in the future, we anticipate that \karmmatwo\ run time would dramatically decrease. For sampling, we use the NUTS \citep{nuts} algorithm which automatically tunes the step size and trajectory length for HMC, thereby reducing the autocorrelation between samples. We perform 100 steps of burn-in followed by 200 samples for each MCMC chain generated in this work. In Appendix \ref{sec:chain_length}, we demonstrate that despite the small number of samples --- particularly given the high dimensionality of our  parameter space --- the posterior distribution of the summary statistics considered is already well sampled by this point.  We attribute this remarkable convergence rate to the fact that wide survey footprints already contain many independent sky patches. Consequently, one realization of the survey area contains within it many nearly-independent samples. This fact boosts the number of ``independent realizations'' used in characterizing the distribution of weak lensing summary statistics, enabling the posterior distributions \textit{of our summary statistics} to be well converged.  Of course, the full posterior of our mass maps is \it not \rm adequately sampled using only 200 mass maps.

\section{Simulation Tests}
\label{sec:sim_tests}

\subsection{Method}

\begin{figure*}
    \centering
    \includegraphics[width=\textwidth]{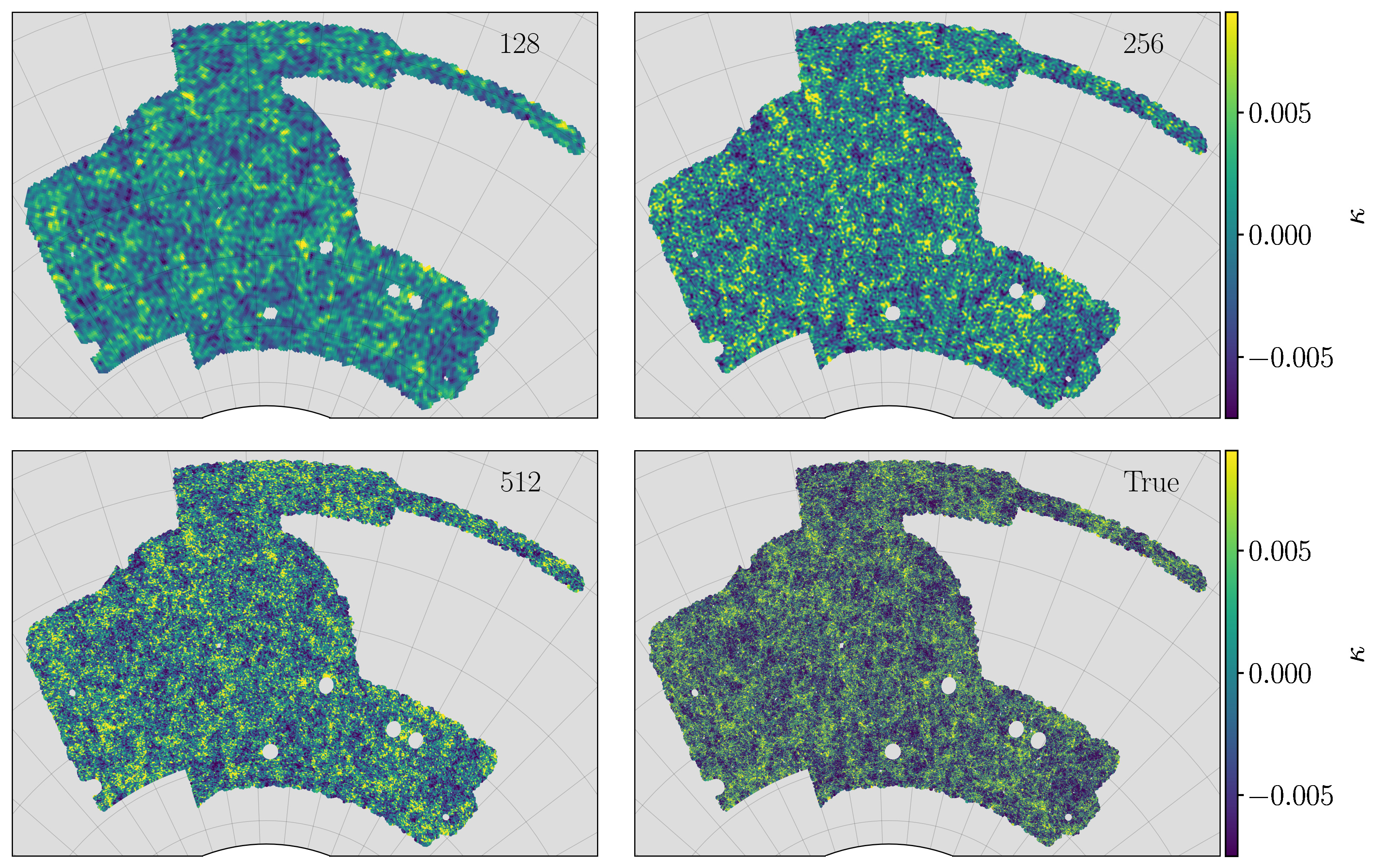}
    \caption[\karmmatwo{} reconstruction visualization.]{Visual comparison of map reconstructions from \karmma{} performed at several resolutions ($N_\mathrm{side}=$ 128, 256, and 512). Each reconstruction is performed for the same mock realization but at a different resolution. For each resolution, we show a randomly chosen convergence map as sampled from the posterior distribution. In the bottom right corner, we show the true convergence map at high resolution ($N_\mathrm{side}=2048$). As we increase the reconstruction resolution, we can see that we begin to recover smaller-scale features of the underlying convergence field. This figure highlights how our updated method can now efficiently forward model the weak lensing field at high resolution.}
    \label{fig:map_comp}
\end{figure*}

We use the suite of 108 full-sky weak lensing simulations generated for the HSC survey in \cite{hsc_mocks}\footnote{\url{http://cosmo.phys.hirosaki-u.ac.jp/takahasi/allsky_raytracing/}} to test the \karmmatwo\ algorithm. The simulations were generated with a fixed flat-$\Lambda$CDM cosmology with cosmological parameters set by the WMAP Y9 \citep{wmap} results: $\Omega_m=0.279$, $h=0.7$, $\sigma_8=0.82$, and $n_s=0.97$. We use the simulations to construct DES Y3-like mocks using the Y3 survey mask, shape noise, and non-tomographic source redshift distribution ($dn/dz$) \citep{y3_shape_cat}. To add shape noise, we first randomly draw the number of galaxies in each pixel as a Poisson realization given the expected number of galaxies.  We then add Gaussian shape noise to the shears with variance $\sigma_{\gamma, i}^2 = \sigma_e^2 / N_i$ where $\sigma_e$ is the shape noise and $N_i$ is the number of source galaxies in pixel $i$. We construct our mock shear maps using the HEALPix \citep{healpix} pixelization scheme of the sphere at HEALPix resolutions of $N_\mathrm{side} = 128$, $256$, and $512$.

For each of the 108 simulations, and each of the three resolutions we considered, we run the \karmmatwo\ algorithm. HEALPix maps suffer aliasing such that only modes up to  $\ell_{\rm max} \approx 2N_{\rm side}$ can be accurately recovered from a map at resolution $N_\mathrm{side}$. For this reason, we reconstruct only the low-pass filtered convergence field up to $\ell_{\rm max}=2N_{\rm side}$. Because of mode-mixing introduced by the exponentiation in equation \ref{eq:lognorm}, if $y$ is band-limited to $\ell_\mathrm{max} = 2 N_\mathrm{side}$, the resulting $\kappa$ field will not be band-limited; instead, is has slightly ($< 1\%$) suppressed power for $\ell$ approaching $\ell_\mathrm{max}$, and a small amount of power at $\ell > \ell_\mathrm{max}$. Removing the excess high $\ell$ power with a low-pass filter is not a good solution: the bias at $\ell \leq \lmax$ remains, while the removed power degrades the resulting 1-point distributions.  Instead, we model the $y$ field up to $\ell = 3 N_\mathrm{side} - 1$ (i.e., $y_{\ell m} = 0$ for $\ell >= 3 N_\mathrm{side} - 1$), and low-pass filter the resulting $\kappa$ map to $\ell_\mathrm{max} = 2 N_\mathrm{side}$. Doing so ensures that the low-pass filtered convergence field has the both the correct power spectrum and one-point distribution. 

\begin{figure*}
\centering
\begin{minipage}[b]{.45\textwidth}
    \includegraphics[width=\columnwidth]{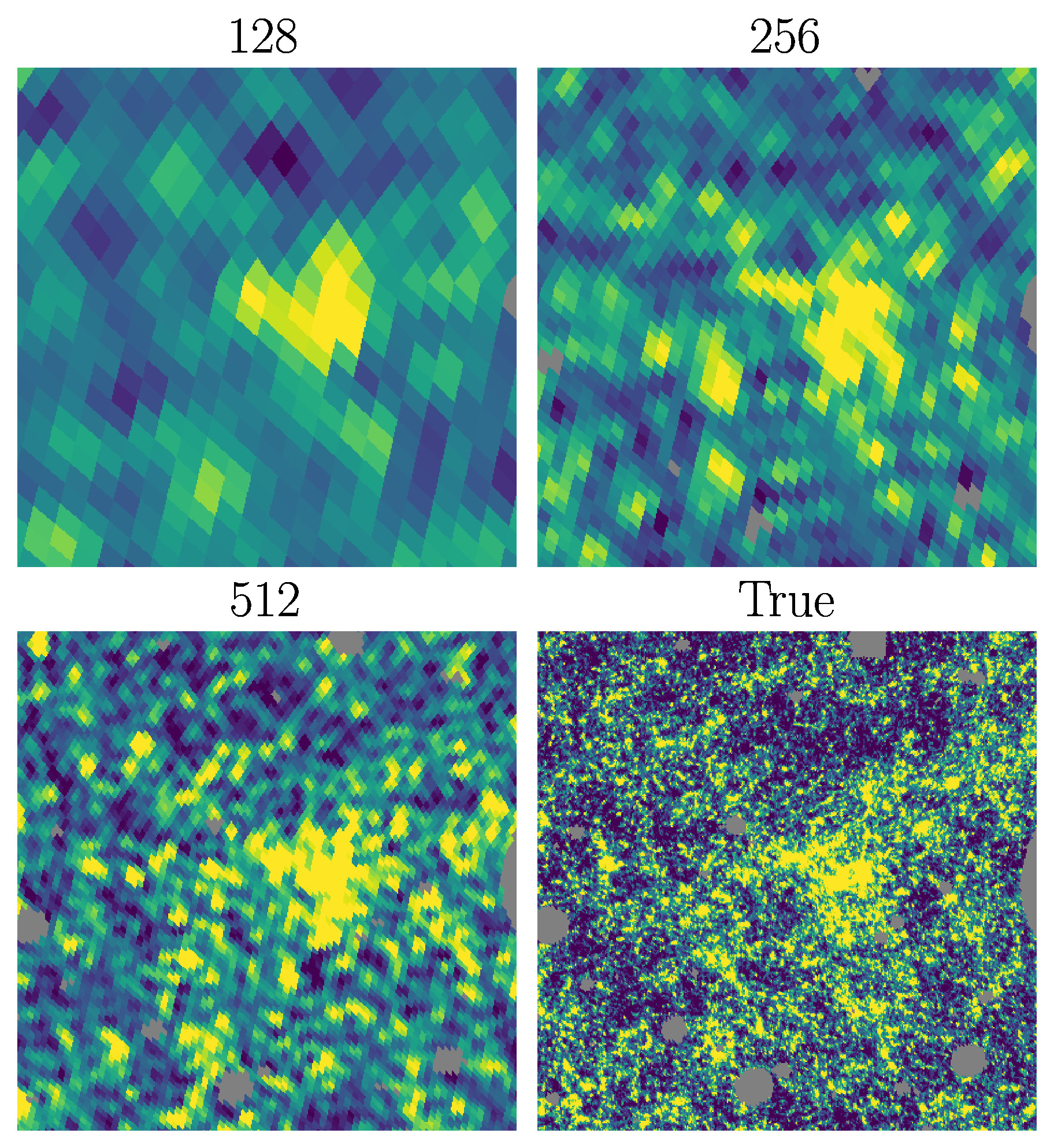}
    \caption[\karmmatwo{} peak visualization.]{Sample \karmmatwo\ maps zoomed in on a 10 deg. $\times$ 10 deg. patch centered around the densest peak in the sky for a randomly chosen simulation.  Each panel corresponds to a different resolution, as labelled.  The bottom-right panel is the true density field sampled at high ($\Nside=2048$) resolution.}
    \label{fig:peak_map}
\end{minipage}\qquad\qquad
\begin{minipage}[b]{.45\textwidth}
    \includegraphics[width=\columnwidth]{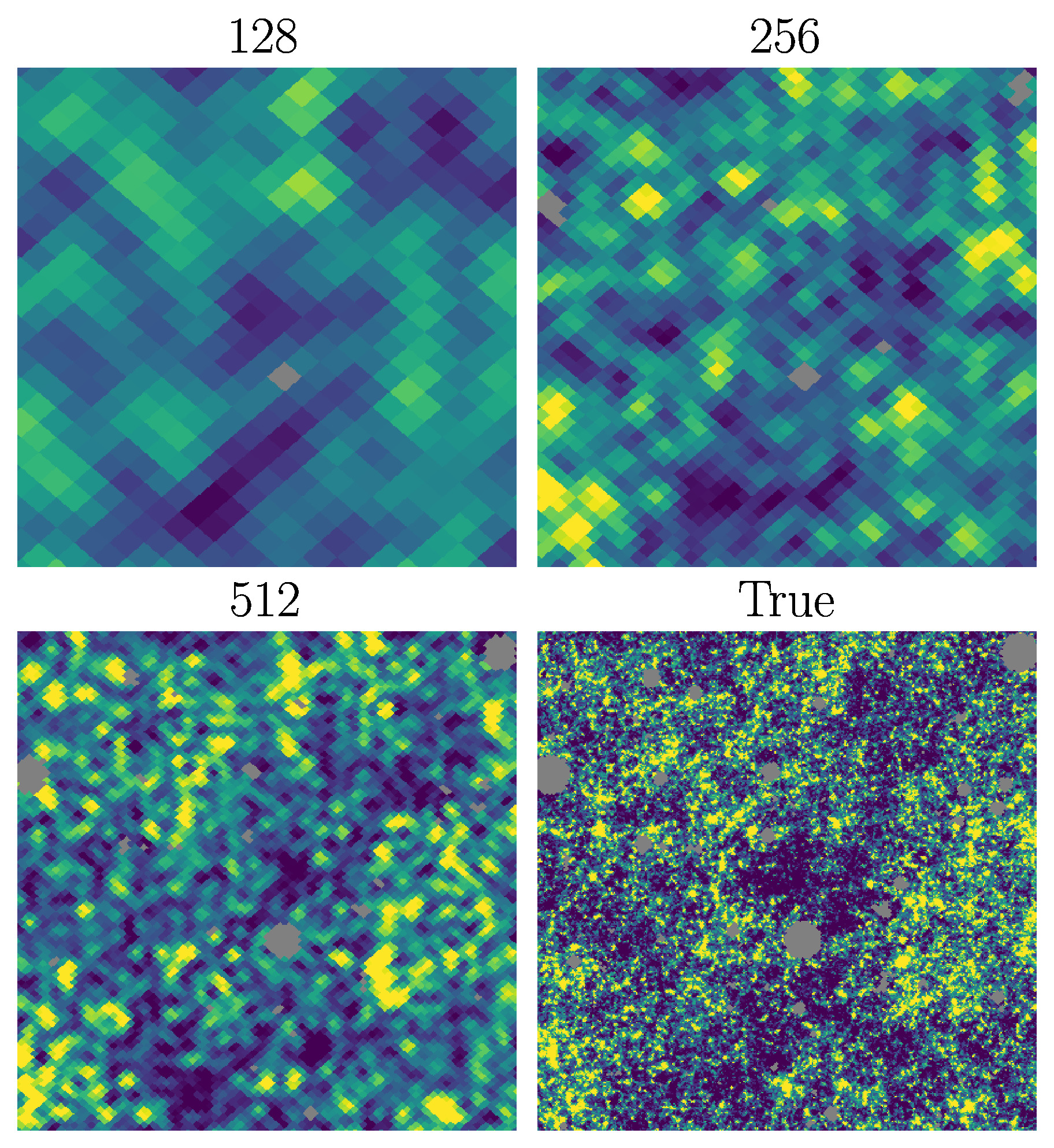}
    \caption[\karmmatwo{} peak visualization.]{Same as Figure~\ref{fig:peak_map}, but now zoomed in on a 10 deg. $\times$ 10 deg. patch centered around the least dense void in the sky.}
    \vspace{0.41in}
    \label{fig:void_map}
\end{minipage}
\end{figure*}

Figure \ref{fig:map_comp} shows a visual comparison between random samples from the \karmmatwo{} posteriors along with the true convergence field. The true map is shown at a very high resolution ($N_\mathrm{side}=2048$, or just under 1~arcmin.) We see that: 1) all the maps have similar large scale structure; and 2) the resolution increases steadily as we move from $\Nside=128$ to $\Nside=512$ without degrading the qualitative aspect of the maps.  This can be better seen in Figures \ref{fig:peak_map} and \ref{fig:void_map}, which zoom into 10~deg $\times$ 10~deg zoom patches centered on: A) the largest density peak in the sky (Fig.~\ref{fig:peak_map}); and B) the least dense void in the sky (Fig.~\ref{fig:void_map}) for one randomly chosen simulation.

We characterize the performance of \karmmatwo\ using a range of summary statistics, namely: 1) the binned power spectrum; 2) the correlation function; 3) the one-point function; 4) peak counts; and 5) void counts. We stress that in these tests we are comparing the summary statistics of our maps marginalized over the full posterior distribution of mass maps, as opposed to calculating the summary statistics of the single maximum-posterior mass map. This is an important point: the maximum a posteriori maps suppress unresolved density fluctuations, giving rise to biased summary statistics.  For further discussion, we refer the reader to \cite{karmma}.

The comparison is performed as follows. For a given cosmology $\vec{\theta}$ and summary-statistic $\vec{s}$, we calculate the mean and covariance matrix of the distribution
\begin{equation}
    P(\vec{s} | \vec{\theta}) = \int P(\vec{s} | \vec{\kappa}) P(\vec{\kappa} | \vec{\theta}) d\vec{\kappa}.
\end{equation}
To do so, we calculate the summary statistics $\vec s$ in each of the maps in the posterior.  We then evaluate the mean and covariance matrix of $\vec s$ across all posterior samples.  The mean summary statistic $\langle \vec s \rangle$ is compared to the mean statistic evaluated from the input simulations.  Because \karmmatwo{} produces low-pass filtered maps, we low-pass filter the simulations at a high resolution and then downgrade to the resolution of the \karmma{} maps before computing the summary statistics in the simulation. The low pass filtering at high resolution is necessary to avoid aliasing in our ``truth'' maps.

\subsection{Results}

\begin{figure*}
    \centering
    \includegraphics[width=\textwidth]{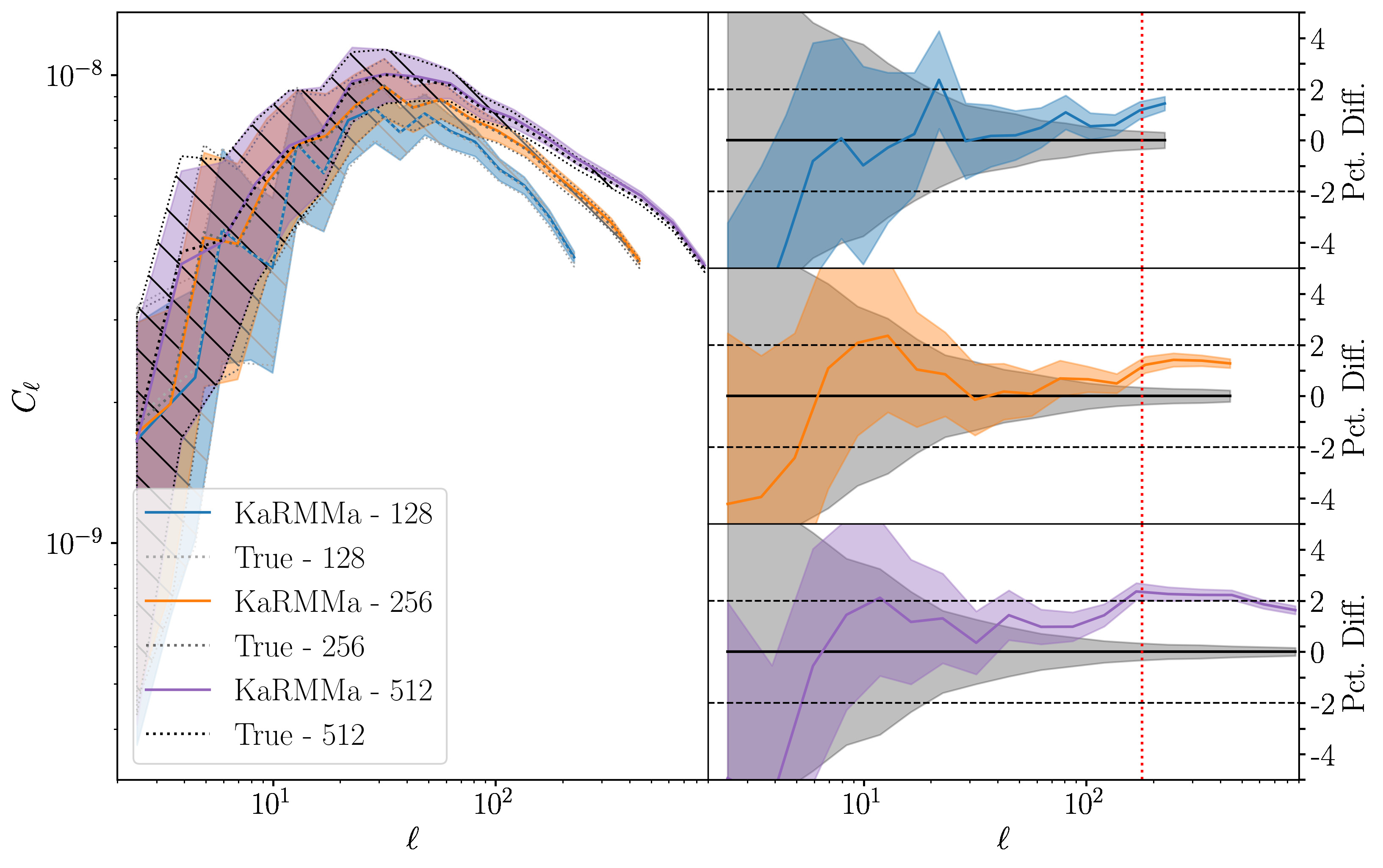}
    \caption[\karmmatwo{} power spectra comparison.]{\textit{Left:} Comparison of the reconstructed power spectrum from \karmma{} samples to the input simulations at multiple resolutions. The bands represent the $1\sigma$ scatter due to cosmic variance. The right panels show the \karmma\ residuals at a resolution of $N_\mathrm{side}=128$ (27~arcmin, \it top),\rm $N_\mathrm{side}=256$ (14~arcmin, \it middle), \rm and $N_\mathrm{side}=512$ (7~arcmin, \it bottom). \rm The bands are the $1 \sigma$ error on the mean.  The horizontal dashed lines are at $\pm$ 2\%, and the red vertical line marks the non-linear scale. As $\ell$ approaches the non-linear scale, the recovered power spectrum becomes biased by $\approx 2\%$ for all resolutions.}
    \label{fig:cl}
\end{figure*}

\begin{figure*}
    \centering
    \includegraphics[width=\textwidth]{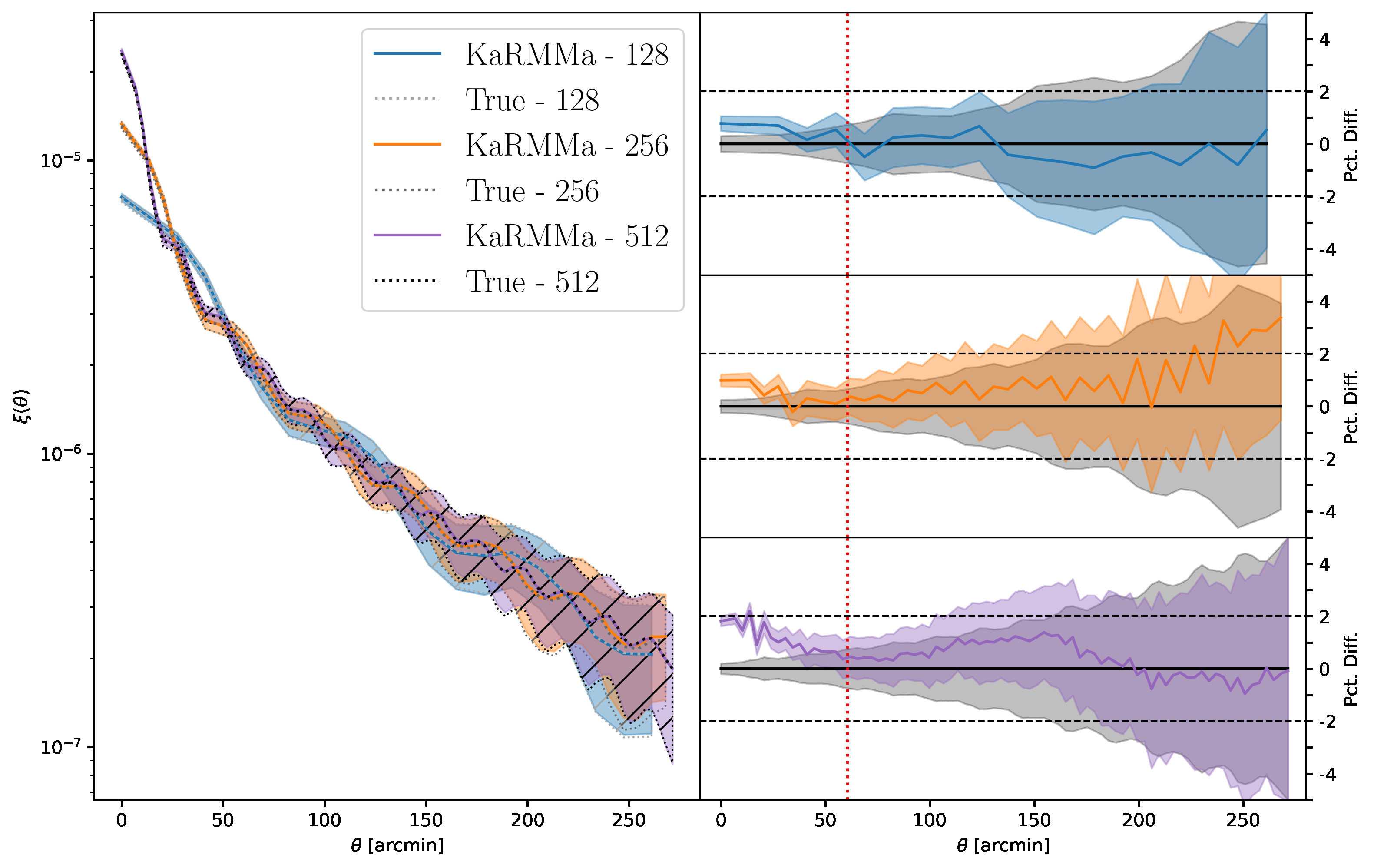}
    \caption[\karmmatwo{} correlation function comparison.]{Comparison of the reconstructed correlation function from \karmma{} to the true correlation function. The layout of this plot is the same as in Figure \ref{fig:cl}. Similarly to the power spectrum, we find that the correlation function from \karmma{} deviates from truth at small scales and worsens as we increase the reconstruction resolution. The vertical line in the right panels corresponds to the non-linear angular scale $\theta_\mathrm{NL}=\pi/\lnl$.}
    \label{fig:xi}
\end{figure*}

\begin{figure*}
    \centering
    \includegraphics[width=\textwidth]{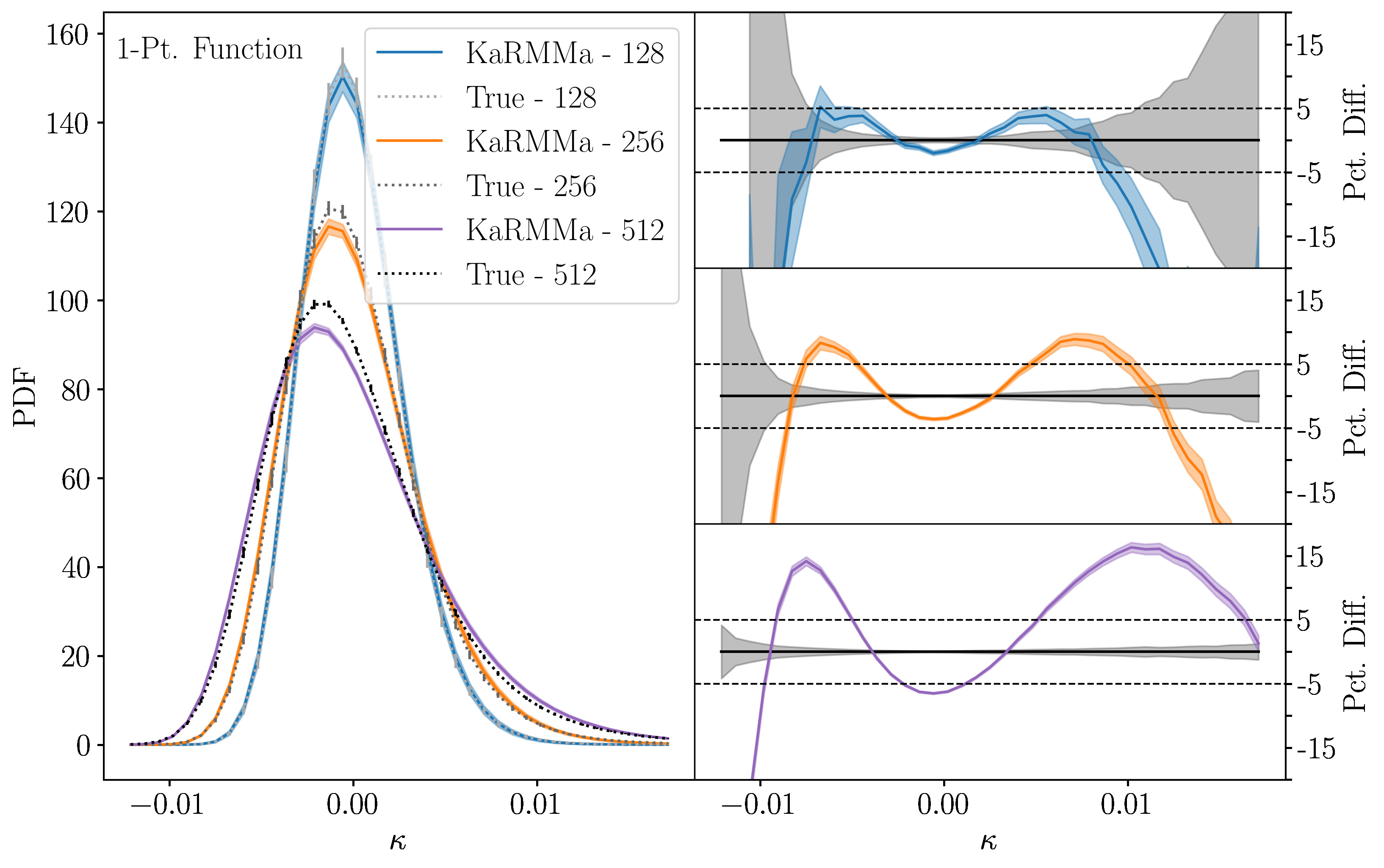}
    \caption[\karmmatwo{} one-point function comparison.]{Comparison of the reconstructed one-point function from \karmma{} to the true one-point function. The layout of this plot is the same as in Figure \ref{fig:cl}. At low resolution, we find that \karmma{} does a good job of recovering the true one-point function at about the 5\% level. At higher resolutions, the one-point function from \karmma{} becomes biased at around the 15\% level, indicating a failure of the lognormal model at small scales.}
    \label{fig:hist}
\end{figure*}

\begin{figure*}
    \centering
    \includegraphics[width=\textwidth]{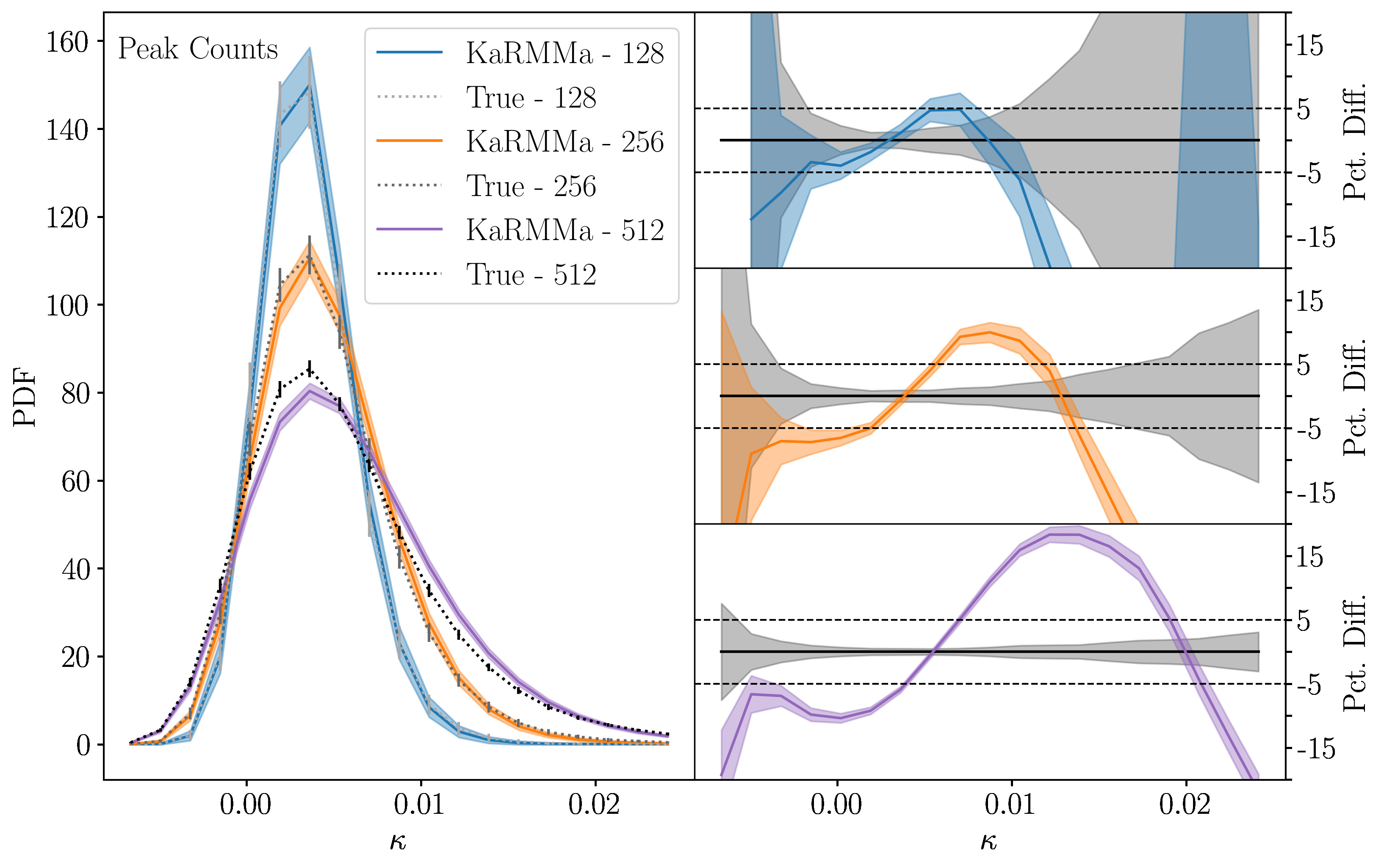}
    \caption[\karmmatwo{} peak counts comparison.]{Comparison of the reconstructed peak counts distribution from \karmma{} to the true peak counts distribution. The layout of this plot is the same as in Figure \ref{fig:cl}. Similarly to the one-point function, we find that \karmma{} does a good job of recovering the correct peak counts distribution at large scales, but fails at the 15\% level or greater at small scales.}
    \label{fig:peak}
\end{figure*}

\begin{figure*}
    \centering
    \includegraphics[width=\textwidth]{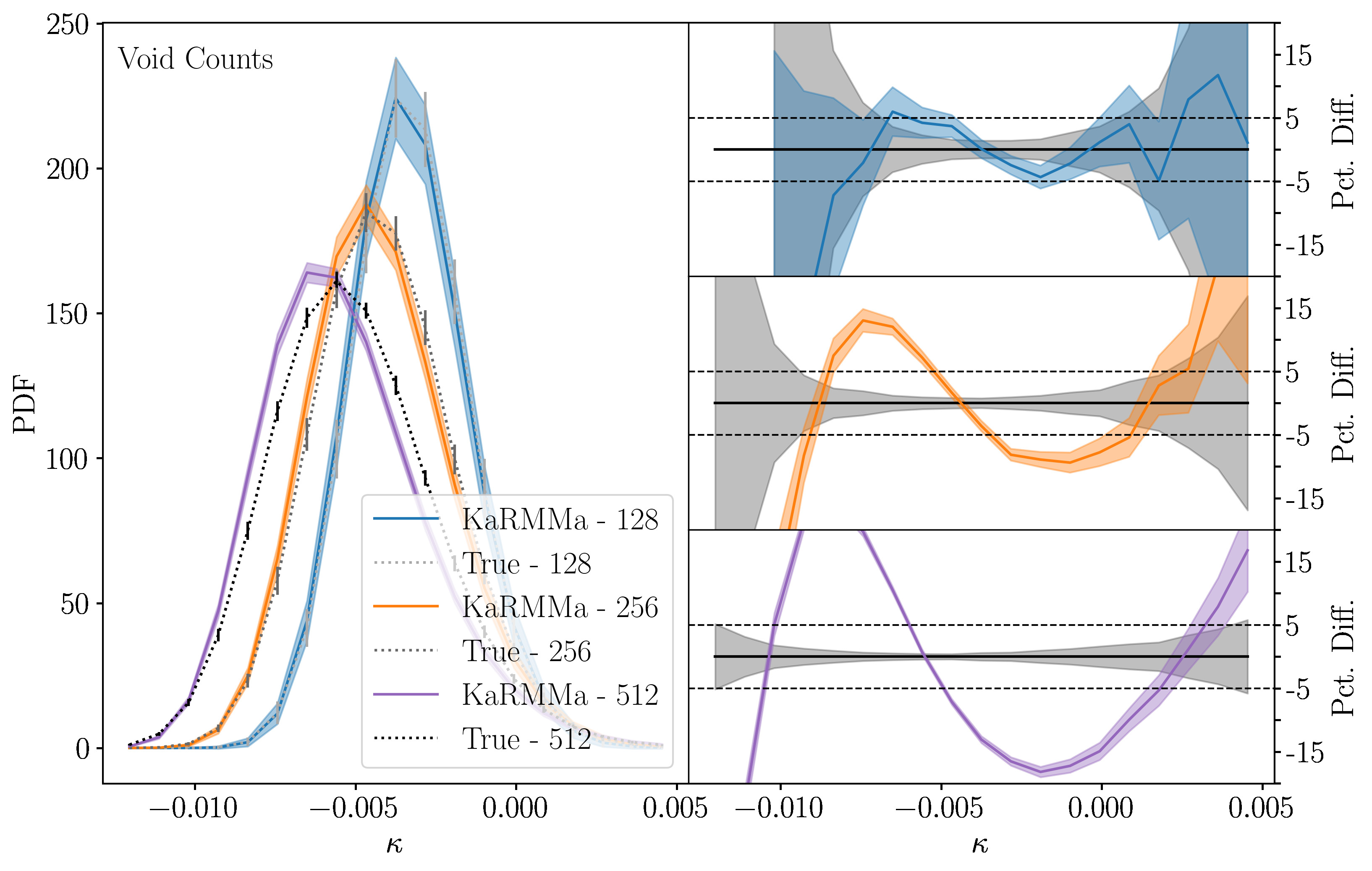}
    \caption[\karmmatwo{} void counts comparison.]{Comparison of the reconstructed void counts distribution from \karmma{} to the true void counts distribution. The layout of this plot is the same as in Figure \ref{fig:cl}. Identically to peak counts, we find that \karmma{} correctly recovers the void counts distribution at low resolution but fails at high resolution.}
    \label{fig:void}
\end{figure*}

Figures \ref{fig:cl} - \ref{fig:void} compare the \karmmatwo\ posteriors for each of the summary statistics to ``truth'' as determined from the simulations.  All plots share a common layout where the left panel overplots the \karmma\ and simulation summary statistics at each of the resolutions tested.  Note the simulation summary statistics change with resolution since the simulation map is always low-pass filtered to $\ell_{\rm max}=2N_{\rm side}$. The bands in the left panel are the $1\sigma$ scatter on the summary statistics as measured using the \karmma\ posterior distributions. The three panels on the right show the residuals for the summary statistics at each of three resolutions we consider. Unlike the left panel, the colored bands here are the \it error on the mean residual \rm as estimated from the 108 independent simulations, i.e. they are smaller than the bands in the left-hand plot by a factor of $\sqrt{108}$.  The black bands centered on zero are the error on the mean due to cosmic variance in the simulations (i.e., they do not involve \karmma{} in any way).

Figure \ref{fig:cl} compares the power spectrum in the simulations and at each resolution. We avoid the prior-dominated parts of the sky by setting the convergence field to zero outside the survey boundary for both simulations and the \karmma\ posteriors.  Since the mask effects are identical, we have not corrected the power spectrum measurements for mask effects. We find that the samples generated by \karmma{} correctly recover the power spectrum at large scales. However, at small non-linear scales the reconstructed power is larger than that in simulations by $\approx$~2\%. Because we are able to make this comparison using 108 simulations, this small excess is detected with high significance.  

The vertical red line in the residual plots is the non-linear scale $\ell_\mathrm{NL}$ defined as follows.  First, given a source at redshift $z$, we set $\lnl(z)=d_A(z)\knl$ with $k_\mathrm{NL}(z)$ predicted by \textsc{PyCCL} as the inverse variance of the displacement field from linear theory. The non-linear scale for a tomographic bin is set by averaging $\lnl(z)$ over the source redshift distribution. The fact that the \karmma{} posterior for the two-point function starts to deviate from truth at $\ell \gtrsim \lnl$ demonstrates the lognormal model fails on strongly non-linear scales.  Indeed, figure~\ref{fig:xi} compares the corresponding correlation function posteriors.  There, we see the bias of the \karmma\ posteriors at zero angular separation increases with increasing resolution. The vertical red line in figure~\ref{fig:xi} corresponds to the non-linear angular scale $\theta_\mathrm{NL} \equiv \pi/\lnl$.

In principle, the recovered power is subject to a second source of bias: aliasing. In real data, the observed shear field is generated by an underlying density field that has power at all scales. When averaging over sky pixels, power in unresolved modes can ``leak'' into the observed angular ranges, leading to an artificial boost in power (aliasing).  This leakage would in turn bias our posteriors.  To test the impact of aliasing, we reran our analyses but modified the input shear field so that it contained no high $\ell$ power. That is, our true underlying density field (and the corresponding shear field) only contains well resolved modes that aren't aliased. Shape noise is then added to the low-pass filtered shear field in each pixel independently (as would be normally) to produce the observed shear field for this test. The posteriors recovered in this case are marginally less biased than in the run subject to aliasing. This test demonstrates that the bias in Figures~\ref{fig:cl} and \ref{fig:xi} is dominated by the failure of the log-normal model on non-linear scales.  Aliasing from high $l$-power in the shear maps is only a minor contaminant in our reconstructions. 

Figure~\ref{fig:hist} compares the posteriors of the 1-point function to simulations. At the lowest resolution of $N_\mathrm{side}=128$ (27 arcmin), we find that the \karmma{} samples recover the one-point function in simulations at an accuracy of $\approx 5\%$ or better.  However, the agreement between the \karmma\ posteriors and simulations becomes increasingly worse with increasing resolution.  At $\Nside=512$ (7 arcmin), the difference in the 1-point function can be as large as 15\%.  Figures~\ref{fig:peak} and \ref{fig:void} comparing the posteriors of the peak and void counts are qualitatively similar to Figure~\ref{fig:hist}: good agreement at low ($\Nside=128$), but increasing differences as resolution improves.

\subsection{Comparison to Other Work}
\label{sec:comparison}

There are many proposals in the literature for how to improve on the traditional Kaiser--Squires algorithm for mass map reconstruction. These broadly fall into two categories: ``best guess'' estimators and fully probabilistic reconstructions.  

``Best guess'' estimators produce a single map which maximizes a specified metric or likelihood function.  This class of methods include:
\begin{enumerate}
    \item Wiener filtering \citep{wiener}: Wiener filtering reconstructs the maximum posterior map under the assumption the convergence field is a Gaussian random field.
    \item \textsc{glimpse} \citep{glimpse_2d, Jeffrey_2018}: \textsc{glimpse} returns the maximum posterior map after adopting a prior of sparsity on the wavelet coefficients of the mass map.
    \item MCALens \citep{starck2021weak}: MCALens combines both the Gaussian and sparsity priors to arrive at a single ``best guess'' map.
    \item DeepMass \citep{Jeffrey_2020}: DeepMass uses machine learning to predict the convergence map by minimizing the mean squared error reconstruction loss.
\end{enumerate}
\cite{y3_mass_map} studied a variety of mass map reconstruction methods, including Wiener filtering and \textsc{glimpse}.  They found these two methods clearly outperform the Kaiser--Squires algorithm when comparing the pixel-level reconstruction quality (Pearson correlation coefficient and mean squared error).  However, they also fail to recover the correct statistical properties of the true mass maps. This failure is generic to all ``best guess'' mass map reconstruction efforts: ``best guess'' maps suppress the amplitude of noise-dominated modes, resulting in a reduction in power at small scales.  For example, at $\ell \sim 500$, the reconstructed power spectrum using either Wiener filtering or \textsc{glimpse} was biased low by more than 80\%.  By contrast, at the same scale, the samples generated by \karmma{} recover the true power spectrum at the level of 2\% error.

Probabilistic reconstruction algorithms --- which includes \karmma{} --- sample the probability distribution $P(\kappa | \gamma_\mathrm{obs})$.  At present, we are aware of three other such algorithms. \cite{Porqueres_2021} approaches this problem by modeling the initial 3D density field of the Universe.  The initial conditions are non-linearly evolved using Lagrangian perturbation theory, and integrated along the line of sight to generate 3D density maps of the Universe.  Their angular resolution is comparable to ours, but are restricted to relatively small patches of sky for which the flat sky approximation is valid.  Based on their Fig. 8, we estimate they recover the true shear power spectrum with an accuracy of $\approx 5\%$, comparable to but slightly worse than \karmma. They do not compare their mass maps to simulations using non-Gaussian summary statistics.

The second algorithm we are aware of is that of \cite{remy2020probabilistic}, who uses machine learning in the form of neural score estimation to sample the posterior distribution of mass maps. Comparing the accuracy of their approach to ours is impossible.  While the $\approx 10\%$ accuracy on the power spectrum reported in that work appears to be worse than that of \karmma, the two algorithms are operating in completely opposite regimes.  Specifically, their simulations are designed to match COSMOS data, which has an area of just $1.64\ {\rm deg}^2$.   Consequently, the largest scales they constrain are roughly the smallest scales we consider.  Their success in recovering the power spectrum at such small scales is a testament to the power of machine learning techniques \citep[see also][]{trenf,cosmogan,kids_gan}, though we caution that the large theoretical uncertainty at such small scales (e.g., due to baryonic processes, \citealt{modelling_baryons}) represent a significant challenge for modeling these scales.  

\cite{alsing_mass_mapping} developed a probabilistic approach for jointly sampling the posterior distribution of shear maps and the shear power spectrum, with the shear modeled as a Gaussian random field. The authors tested their reconstruction on a $100\ \mathrm{deg}^2$ mock survey at an angular resolution of roughly 5~arcmin using the flat sky approximation. The power spectrum of their reconstructed shear maps recovers the power spectrum of the true shear field to within the uncertainty of their posterior distribution. However, as they do not show power spectrum residuals, it is difficult to compare the accuracy of their reconstruction to ours.  We note that in \citet{karmma} we founds that adopting a Gaussian prior for the convergence field enables one to recover maps that are of comparable quality to \karmma\ in terms of the accuracy of the resulting power spectra.  However, the recovered non-Gaussian summary statistics exhibit a larger bias than those obtained using \karmma.

In summary, the statistical properties of mass maps are much better reproduced by probabilistic sampling algorithms than by ``best guess'' maps.  Compared to the other three probabilistic algorithms currently available in the literature, \karmma\ achieves either comparable or somewhat better accuracy in terms of the reconstructed power spectrum.  We have also, for the first time, quantitatively characterized the performance of a probabilistic algorithm using non-Gaussian statistics. At the time of writing, \karmma\ is the only such algorithm capable of operating on large survey areas for which the flat sky approximation is invalid.

\section{Summary and Conclusions}
\label{sec:conclusion}

We have presented an updated version of \karmma, an algorithm for mass map reconstruction. \karmma\ forward models the observed shear field subject to a lognormal prior. By reparameterizing the problem in terms of the spherical harmonic coefficients of the transformed convergence field (i.e., $y=\ln(\kappa+\lambda)$), we are able to decorrelate our model parameters in the prior.  This feature, in combination with the existence of efficient spherical harmonic transformations enables us to improve the scaling of our algorithm to efficiently perform reconstruction at resolutions that were formerly computationally intractable. As with our first iteration of the algorithm \citep{karmma}, \karmmatwo{} is fully Bayesian and generates a library of maps that sample the posterior distribution of maps consistent with the observed shear field.

Figures~\ref{fig:cl} through \ref{fig:void} summarize our results. The figures compare the \karmmatwo\ posteriors to simulations for a variety of summary statistics: power spectrum, correlation function, 1-point function, and peak and void counts.  In all cases, \karmmatwo\ returns accurate (5\% or better) posteriors for all statistics at $\Nside=128$ resolution.  As resolution increases, the accuracy of the posteriors decreases, with systematic differences in non-Gaussian summary statistics reaching as high as $\approx 15\%$ at $\Nside=512$.

Given that \karmmatwo\ can now return high resolution posteriors, the main difficulty moving forward is our ability to resolve strongly non-linear scales.  At these scales, the lognormal approximation fails to be adequate, driving the biases observed in Figures~\ref{fig:cl} through \ref{fig:void}. To overcome this limitation, we need to improve upon the lognormal approximation.  Ideally, one would replace the lognormal model with full gravity simulations.  Doing so is nontrivial, and remains an active area of research \citep{Jasche_2013, borg, Porqueres_2021}. Moving forward, we intend instead to augment the lognormal model with machine-learning techniques that enable us to accurately reconstruct the true non-linear density field (Fiedorowicz et al. in preparation).  Future releases of \karmma\ will also enable: 1) tomographic mass map reconstruction \citep{supranta}; and 2) marginalization over cosmological parameters (Boruah et al., in preparation).

Despite its current limitations, we consider \karmmatwo\ an unquestionable success.  First, even at $\Nside=512$, the $\lesssim 15\%$ biases in our posteriors are a dramatic improvement relative to other mass mapping techniques available in the literature (see Section \ref{sec:comparison}).  Second, our algorithm is one of the few approaches that works in a spherical sky.  Third, our algorithm is fully Bayesian, returning proper posteriors as opposed to a single ``best guess'' mass map.  This last point is particularly important, since ``best guess'' maps are necessarily strongly biased \citep[see][]{karmma}.  For all these reasons, \karmmatwo\ represents a significant step forward in our ability to reconstruct the matter density field of the Universe from weak lensing data.

Upon publication, the updated version of the \karmma{} code will be made publicly available on GitHub at \url{https://github.com/pierfied/karmma}.

\section*{Acknowledgements}

ER and PF are supported by NSF grant 2009401.  ER also receives support from DOE grant DE-SC0009913.

We would like to thank \citet{Takahashi_2017} and collaborators for making their full-sky weak lensing simulations publicly available.  Without them, this work would not have been possible.

%%%%%%%%%%%%%%%%%%%%%%%%%%%%%%%%%%%%%%%%%%%%%%%%%%
\section*{Data Availability}

The data underlying this article was derived from \cite{Takahashi_2017} which can be accessed at: \url{http://cosmo.phys.hirosaki-u.ac.jp/takahasi/allskyraytracing/}. The derived data generated in this research will be shared on reasonable request to the corresponding author.



\bibliographystyle{mnras}
\bibliography{karmma_2.0} % if your bibtex file is called example.bib

\newcommand{\noop}[1]{}
\begin{thebibliography}{}
\makeatletter
\relax
\def\mn@urlcharsother{\let\do\@makeother \do\$\do\&\do\#\do\^\do\_\do\%\do\~}
\def\mn@doi{\begingroup\mn@urlcharsother \@ifnextchar [ {\mn@doi@}
  {\mn@doi@[]}}
\def\mn@doi@[#1]#2{\def\@tempa{#1}\ifx\@tempa\@empty \href
  {http://dx.doi.org/#2} {doi:#2}\else \href {http://dx.doi.org/#2} {#1}\fi
  \endgroup}
\def\mn@eprint#1#2{\mn@eprint@#1:#2::\@nil}
\def\mn@eprint@arXiv#1{\href {http://arxiv.org/abs/#1} {{\tt arXiv:#1}}}
\def\mn@eprint@dblp#1{\href {http://dblp.uni-trier.de/rec/bibtex/#1.xml}
  {dblp:#1}}
\def\mn@eprint@#1:#2:#3:#4\@nil{\def\@tempa {#1}\def\@tempb {#2}\def\@tempc
  {#3}\ifx \@tempc \@empty \let \@tempc \@tempb \let \@tempb \@tempa \fi \ifx
  \@tempb \@empty \def\@tempb {arXiv}\fi \@ifundefined
  {mn@eprint@\@tempb}{\@tempb:\@tempc}{\expandafter \expandafter \csname
  mn@eprint@\@tempb\endcsname \expandafter{\@tempc}}}

\bibitem[\protect\citeauthoryear{Abbott et~al.,}{Abbott
  et~al.}{2022}]{y3_cosmology}
Abbott T.,  et~al., 2022, \mn@doi [Physical Review D]
  {10.1103/physrevd.105.023520}, 105

\bibitem[\protect\citeauthoryear{Alsing, Heavens, Jaffe, Kiessling, Wandelt  \&
  Hoffmann}{Alsing et~al.}{2015}]{alsing_mass_mapping}
Alsing J.,  Heavens A.,  Jaffe A.~H.,  Kiessling A.,  Wandelt B.,   Hoffmann
  T.,  2015, \mn@doi [Monthly Notices of the Royal Astronomical Society]
  {10.1093/mnras/stv2501}, 455, 4452

\bibitem[\protect\citeauthoryear{{Alsing}, {Heavens}  \& {Jaffe}}{{Alsing}
  et~al.}{2017}]{alsingetal17}
{Alsing} J.,  {Heavens} A.,   {Jaffe} A.~H.,  2017, \mn@doi [\mnras]
  {10.1093/mnras/stw3161}, \href
  {https://ui.adsabs.harvard.edu/abs/2017MNRAS.466.3272A} {466, 3272}

\bibitem[\protect\citeauthoryear{Bingham et~al.,}{Bingham et~al.}{2019}]{pyro}
Bingham E.,  et~al., 2019, J. Mach. Learn. Res., 20, 973–978

\bibitem[\protect\citeauthoryear{Blas, Lesgourgues  \& Tram}{Blas
  et~al.}{2011}]{Blas_2011}
Blas D.,  Lesgourgues J.,   Tram T.,  2011, \mn@doi [Journal of Cosmology and
  Astroparticle Physics] {10.1088/1475-7516/2011/07/034}, 2011, 034–034

\bibitem[\protect\citeauthoryear{Boruah, Rozo  \& Fiedorowicz}{Boruah
  et~al.}{2022}]{supranta}
Boruah S.~S.,  Rozo E.,   Fiedorowicz P.,  2022, Map-based cosmology inference
  with lognormal cosmic shear maps, \mn@doi{10.48550/ARXIV.2204.13216}, \url
  {https://arxiv.org/abs/2204.13216}

\bibitem[\protect\citeauthoryear{{Castro}, {Heavens}  \& {Kitching}}{{Castro}
  et~al.}{2005}]{spherical_KS}
{Castro} P.~G.,  {Heavens} A.~F.,   {Kitching} T.~D.,  2005, \mn@doi [\prd]
  {10.1103/PhysRevD.72.023516}, \href
  {https://ui.adsabs.harvard.edu/abs/2005PhRvD..72b3516C} {72, 023516}

\bibitem[\protect\citeauthoryear{Cheng, Ting, M{\'{e}}nard  \& Bruna}{Cheng
  et~al.}{2020}]{scattering_transforms}
Cheng S.,  Ting Y.-S.,  M{\'{e}}nard B.,   Bruna J.,  2020, \mn@doi [Monthly
  Notices of the Royal Astronomical Society] {10.1093/mnras/staa3165}, 499,
  5902

\bibitem[\protect\citeauthoryear{Chisari et~al.,}{Chisari
  et~al.}{2019a}]{modelling_baryons}
Chisari N.~E.,  et~al., 2019a, \mn@doi [The Open Journal of Astrophysics]
  {10.21105/astro.1905.06082}, 2

\bibitem[\protect\citeauthoryear{Chisari et~al.,}{Chisari et~al.}{2019b}]{ccl}
Chisari N.~E.,  et~al., 2019b, \mn@doi [The Astrophysical Journal Supplement
  Series] {10.3847/1538-4365/ab1658}, 242, 2

\bibitem[\protect\citeauthoryear{Clerkin et~al.,}{Clerkin
  et~al.}{2016}]{des_lognorm}
Clerkin L.,  et~al., 2016, \mn@doi [Monthly Notices of the Royal Astronomical
  Society] {10.1093/mnras/stw2106}, 466, 1444

\bibitem[\protect\citeauthoryear{Dai \& Seljak}{Dai \& Seljak}{2022}]{trenf}
Dai B.,  Seljak U.,  2022, Translation and Rotation Equivariant Normalizing
  Flow (TRENF) for Optimal Cosmological Analysis,
  \mn@doi{10.48550/ARXIV.2202.05282}, \url {https://arxiv.org/abs/2202.05282}

\bibitem[\protect\citeauthoryear{Dietrich \& Hartlap}{Dietrich \&
  Hartlap}{2010}]{Dietrich_2010}
Dietrich J.~P.,  Hartlap J.,  2010, \mn@doi [Monthly Notices of the Royal
  Astronomical Society] {10.1111/j.1365-2966.2009.15948.x}, 402, 1049–1058

\bibitem[\protect\citeauthoryear{Fiedorowicz, Rozo, Boruah, Chang  \&
  Gatti}{Fiedorowicz et~al.}{2022}]{karmma}
Fiedorowicz P.,  Rozo E.,  Boruah S.~S.,  Chang C.,   Gatti M.,  2022, \mn@doi
  [Monthly Notices of the Royal Astronomical Society] {10.1093/mnras/stac468},
  512, 73

\bibitem[\protect\citeauthoryear{Fluri, Kacprzak, Refregier, Amara, Lucchi  \&
  Hofmann}{Fluri et~al.}{2018}]{Fluri_2018}
Fluri J.,  Kacprzak T.,  Refregier A.,  Amara A.,  Lucchi A.,   Hofmann T.,
  2018, \mn@doi [Physical Review D] {10.1103/physrevd.98.123518}, 98

\bibitem[\protect\citeauthoryear{Friedrich, Uhlemann, Villaescusa-Navarro,
  Baldauf, Manera  \& Nishimichi}{Friedrich et~al.}{2020}]{Friedrich_2020}
Friedrich O.,  Uhlemann C.,  Villaescusa-Navarro F.,  Baldauf T.,  Manera M.,
  Nishimichi T.,  2020, \mn@doi [Monthly Notices of the Royal Astronomical
  Society] {10.1093/mnras/staa2160}, 498, 464–483

\bibitem[\protect\citeauthoryear{Fu et~al.,}{Fu et~al.}{2014}]{fu_bispectrum}
Fu L.,  et~al., 2014, \mn@doi [Monthly Notices of the Royal Astronomical
  Society] {10.1093/mnras/stu754}, 441, 2725

\bibitem[\protect\citeauthoryear{Gatti et~al.,}{Gatti
  et~al.}{2020}]{Gatti_2020}
Gatti M.,  et~al., 2020, \mn@doi [Monthly Notices of the Royal Astronomical
  Society] {10.1093/mnras/staa2680}, 498, 4060–4087

\bibitem[\protect\citeauthoryear{Gatti et~al.,}{Gatti
  et~al.}{2021}]{y3_shape_cat}
Gatti M.,  et~al., 2021, \mn@doi [Monthly Notices of the Royal Astronomical
  Society] {10.1093/mnras/stab918}, 504, 4312

\bibitem[\protect\citeauthoryear{Gorski, Hivon, Banday, Wandelt, Hansen,
  Reinecke  \& Bartelmann}{Gorski et~al.}{2005}]{healpix}
Gorski K.~M.,  Hivon E.,  Banday A.~J.,  Wandelt B.~D.,  Hansen F.~K.,
  Reinecke M.,   Bartelmann M.,  2005, \mn@doi [The Astrophysical Journal]
  {10.1086/427976}, 622, 759

\bibitem[\protect\citeauthoryear{Gupta, Matilla, Hsu  \& Haiman}{Gupta
  et~al.}{2018}]{Gupta_2018}
Gupta A.,  Matilla J. M.~Z.,  Hsu D.,   Haiman Z.,  2018, \mn@doi [Physical
  Review D] {10.1103/physrevd.97.103515}, 97

\bibitem[\protect\citeauthoryear{Heymans et~al.,}{Heymans
  et~al.}{2021}]{Heymans_2021}
Heymans C.,  et~al., 2021, \mn@doi [Astronomy & Astrophysics]
  {10.1051/0004-6361/202039063}, 646, A140

\bibitem[\protect\citeauthoryear{{Hikage} et~al.,}{{Hikage}
  et~al.}{2019}]{hikage_etal19}
{Hikage} C.,  et~al., 2019, \mn@doi [\pasj] {10.1093/pasj/psz010}, \href
  {https://ui.adsabs.harvard.edu/abs/2019PASJ...71...43H} {71, 43}

\bibitem[\protect\citeauthoryear{Hinshaw et~al.,}{Hinshaw et~al.}{2013}]{wmap}
Hinshaw G.,  et~al., 2013, \mn@doi [The Astrophysical Journal Supplement
  Series] {10.1088/0067-0049/208/2/19}, 208, 19

\bibitem[\protect\citeauthoryear{Hoffman \& Gelman}{Hoffman \&
  Gelman}{2011}]{nuts}
Hoffman M.~D.,  Gelman A.,  2011, The No-U-Turn Sampler: Adaptively Setting
  Path Lengths in Hamiltonian Monte Carlo, \mn@doi{10.48550/ARXIV.1111.4246},
  \url {https://arxiv.org/abs/1111.4246}

\bibitem[\protect\citeauthoryear{Jasche \& Wandelt}{Jasche \&
  Wandelt}{2013}]{Jasche_2013}
Jasche J.,  Wandelt B.~D.,  2013, \mn@doi [Monthly Notices of the Royal
  Astronomical Society] {10.1093/mnras/stt449}, 432, 894–913

\bibitem[\protect\citeauthoryear{Jeffrey et~al.,}{Jeffrey
  et~al.}{2018}]{Jeffrey_2018}
Jeffrey N.,  et~al., 2018, \mn@doi [Monthly Notices of the Royal Astronomical
  Society] {10.1093/mnras/sty1252}, 479, 2871–2888

\bibitem[\protect\citeauthoryear{Jeffrey, Lanusse, Lahav  \& Starck}{Jeffrey
  et~al.}{2020a}]{Jeffrey_2020}
Jeffrey N.,  Lanusse F.,  Lahav O.,   Starck J.-L.,  2020a, \mn@doi [Monthly
  Notices of the Royal Astronomical Society] {10.1093/mnras/staa127}, 492,
  5023–5029

\bibitem[\protect\citeauthoryear{Jeffrey, Alsing  \& Lanusse}{Jeffrey
  et~al.}{2020b}]{Jeffrey_2020_inference}
Jeffrey N.,  Alsing J.,   Lanusse F.,  2020b, \mn@doi [Monthly Notices of the
  Royal Astronomical Society] {10.1093/mnras/staa3594}, 501, 954–969

\bibitem[\protect\citeauthoryear{Jeffrey et~al.,}{Jeffrey
  et~al.}{2021}]{y3_mass_map}
Jeffrey N.,  et~al., 2021, \mn@doi [Monthly Notices of the Royal Astronomical
  Society] {10.1093/mnras/stab1495}

\bibitem[\protect\citeauthoryear{Jung, Namikawa, Liguori, Munshi  \&
  Heavens}{Jung et~al.}{2021}]{jung2021integrated}
Jung G.,  Namikawa T.,  Liguori M.,  Munshi D.,   Heavens A.,  2021, The
  integrated angular bispectrum of weak lensing (\mn@eprint {arXiv}
  {2102.05521})

\bibitem[\protect\citeauthoryear{{Kaiser} \& {Squires}}{{Kaiser} \&
  {Squires}}{1993}]{KaiserSquires}
{Kaiser} N.,  {Squires} G.,  1993, \mn@doi [\apj] {10.1086/172297}, \href
  {https://ui.adsabs.harvard.edu/abs/1993ApJ...404..441K} {404, 441}

\bibitem[\protect\citeauthoryear{Kratochvil, Haiman  \& May}{Kratochvil
  et~al.}{2010}]{Kratochvil_2010}
Kratochvil J.~M.,  Haiman Z.,   May M.,  2010, \mn@doi [Physical Review D]
  {10.1103/physrevd.81.043519}, 81

\bibitem[\protect\citeauthoryear{Kratochvil, Lim, Wang, Haiman, May  \&
  Huffenberger}{Kratochvil et~al.}{2012}]{Kratochvil_2012}
Kratochvil J.~M.,  Lim E.~A.,  Wang S.,  Haiman Z.,  May M.,   Huffenberger K.,
   2012, \mn@doi [Physical Review D] {10.1103/physrevd.85.103513}, 85

\bibitem[\protect\citeauthoryear{{Lanusse}, {Starck}, {Leonard}  \&
  {Pires}}{{Lanusse} et~al.}{2016}]{glimpse_2d}
{Lanusse} F.,  {Starck} J.~L.,  {Leonard} A.,   {Pires} S.,  2016, \mn@doi
  [\aap] {10.1051/0004-6361/201628278}, \href
  {https://ui.adsabs.harvard.edu/abs/2016A&A...591A...2L} {591, A2}

\bibitem[\protect\citeauthoryear{Lewis, Challinor  \& Lasenby}{Lewis
  et~al.}{2000}]{Lewis_2000}
Lewis A.,  Challinor A.,   Lasenby A.,  2000, \mn@doi [The Astrophysical
  Journal] {10.1086/309179}, 538, 473–476

\bibitem[\protect\citeauthoryear{Mustafa, Bard, Bhimji, Luki{\'{c}}, Al-Rfou
  \& Kratochvil}{Mustafa et~al.}{2019}]{cosmogan}
Mustafa M.,  Bard D.,  Bhimji W.,  Luki{\'{c}} Z.,  Al-Rfou R.,   Kratochvil
  J.~M.,  2019, \mn@doi [Computational Astrophysics and Cosmology]
  {10.1186/s40668-019-0029-9}, 6

\bibitem[\protect\citeauthoryear{Neal}{Neal}{2012}]{neal2012mcmc}
Neal R.~M.,  2012, MCMC using Hamiltonian dynamics (\mn@eprint {arXiv}
  {1206.1901})

\bibitem[\protect\citeauthoryear{Oguri et~al.,}{Oguri
  et~al.}{2017}]{Oguri_2017}
Oguri M.,  et~al., 2017, \mn@doi [Publications of the Astronomical Society of
  Japan] {10.1093/pasj/psx070}, 70

\bibitem[\protect\citeauthoryear{Peel, Lin, Lanusse, Leonard, Starck  \&
  Kilbinger}{Peel et~al.}{2017}]{Peel_2017}
Peel A.,  Lin C.-A.,  Lanusse F.,  Leonard A.,  Starck J.-L.,   Kilbinger M.,
  2017, \mn@doi [Astronomy & Astrophysics] {10.1051/0004-6361/201629928}, 599,
  A79

\bibitem[\protect\citeauthoryear{Petri, Haiman, Hui, May  \& Kratochvil}{Petri
  et~al.}{2013}]{Petri_2013}
Petri A.,  Haiman Z.,  Hui L.,  May M.,   Kratochvil J.~M.,  2013, \mn@doi
  [Physical Review D] {10.1103/physrevd.88.123002}, 88

\bibitem[\protect\citeauthoryear{{Porqueres}, {Kodi Ramanah}, {Jasche}  \&
  {Lavaux}}{{Porqueres} et~al.}{2019}]{fbi_foreground_contamination}
{Porqueres} N.,  {Kodi Ramanah} D.,  {Jasche} J.,   {Lavaux} G.,  2019, \mn@doi
  [\aap] {10.1051/0004-6361/201834844}, \href
  {https://ui.adsabs.harvard.edu/abs/2019A&A...624A.115P} {624, A115}

\bibitem[\protect\citeauthoryear{Porqueres, Heavens, Mortlock  \&
  Lavaux}{Porqueres et~al.}{2021a}]{Porqueres_2021}
Porqueres N.,  Heavens A.,  Mortlock D.,   Lavaux G.,  2021a, \mn@doi [Monthly
  Notices of the Royal Astronomical Society] {10.1093/mnras/stab204}, 502,
  3035–3044

\bibitem[\protect\citeauthoryear{Porqueres, Heavens, Mortlock  \&
  Lavaux}{Porqueres et~al.}{2021b}]{borg}
Porqueres N.,  Heavens A.,  Mortlock D.,   Lavaux G.,  2021b, \mn@doi [Monthly
  Notices of the Royal Astronomical Society] {10.1093/mnras/stab3234}, 509,
  3194

\bibitem[\protect\citeauthoryear{{Pyne} \& {Joachimi}}{{Pyne} \&
  {Joachimi}}{2021}]{pyne_joachimi_21}
{Pyne} S.,  {Joachimi} B.,  2021, \mn@doi [\mnras] {10.1093/mnras/stab413},
  \href {https://ui.adsabs.harvard.edu/abs/2021MNRAS.503.2300P} {503, 2300}

\bibitem[\protect\citeauthoryear{Remy, Lanusse, Ramzi, Liu, Jeffrey  \&
  Starck}{Remy et~al.}{2020}]{remy2020probabilistic}
Remy B.,  Lanusse F.,  Ramzi Z.,  Liu J.,  Jeffrey N.,   Starck J.-L.,  2020,
  Probabilistic Mapping of Dark Matter by Neural Score Matching (\mn@eprint
  {arXiv} {2011.08271})

\bibitem[\protect\citeauthoryear{{Ribli}, {Pataki}, {Zorrilla Matilla}, {Hsu},
  {Haiman}  \& {Csabai}}{{Ribli} et~al.}{2019}]{Ribli_2019}
{Ribli} D.,  {Pataki} B.~{\'A}.,  {Zorrilla Matilla} J.~M.,  {Hsu} D.,
  {Haiman} Z.,   {Csabai} I.,  2019, \mn@doi [\mnras] {10.1093/mnras/stz2610},
  \href {https://ui.adsabs.harvard.edu/abs/2019MNRAS.490.1843R} {490, 1843}

\bibitem[\protect\citeauthoryear{Shan et~al.,}{Shan et~al.}{2017}]{Shan_2017}
Shan H.,  et~al., 2017, \mn@doi [Monthly Notices of the Royal Astronomical
  Society] {10.1093/mnras/stx2837}, 474, 1116–1134

\bibitem[\protect\citeauthoryear{Smith et~al.,}{Smith
  et~al.}{2003}]{halofit_orig}
Smith R.~E.,  et~al., 2003, \mn@doi [Monthly Notices of the Royal Astronomical
  Society] {10.1046/j.1365-8711.2003.06503.x}, 341, 1311

\bibitem[\protect\citeauthoryear{Starck, Themelis, Jeffrey, Peel  \&
  Lanusse}{Starck et~al.}{2021}]{starck2021weak}
Starck J.~L.,  Themelis K.~E.,  Jeffrey N.,  Peel A.,   Lanusse F.,  2021, Weak
  lensing mass reconstruction using sparsity and a Gaussian random field
  (\mn@eprint {arXiv} {2102.04127})

\bibitem[\protect\citeauthoryear{Takada \& Jain}{Takada \&
  Jain}{2004}]{takada_bispectrum}
Takada M.,  Jain B.,  2004, \mn@doi [Monthly Notices of the Royal Astronomical
  Society] {10.1111/j.1365-2966.2004.07410.x}, 348, 897

\bibitem[\protect\citeauthoryear{Takahashi, Sato, Nishimichi, Taruya  \&
  Oguri}{Takahashi et~al.}{2012}]{halofit_revised}
Takahashi R.,  Sato M.,  Nishimichi T.,  Taruya A.,   Oguri M.,  2012, \mn@doi
  [The Astrophysical Journal] {10.1088/0004-637x/761/2/152}, 761, 152

\bibitem[\protect\citeauthoryear{Takahashi, Hamana, Shirasaki, Namikawa,
  Nishimichi, Osato  \& Shiroyama}{Takahashi et~al.}{2017a}]{hsc_mocks}
Takahashi R.,  Hamana T.,  Shirasaki M.,  Namikawa T.,  Nishimichi T.,  Osato
  K.,   Shiroyama K.,  2017a, \mn@doi [The Astrophysical Journal]
  {10.3847/1538-4357/aa943d}, 850, 24

\bibitem[\protect\citeauthoryear{Takahashi, Hamana, Shirasaki, Namikawa,
  Nishimichi, Osato  \& Shiroyama}{Takahashi et~al.}{2017b}]{Takahashi_2017}
Takahashi R.,  Hamana T.,  Shirasaki M.,  Namikawa T.,  Nishimichi T.,  Osato
  K.,   Shiroyama K.,  2017b, \mn@doi [The Astrophysical Journal]
  {10.3847/1538-4357/aa943d}, 850, 24

\bibitem[\protect\citeauthoryear{Taruya, Takada, Hamana, Kayo  \&
  Futamase}{Taruya et~al.}{2002}]{Taruya_2002}
Taruya A.,  Takada M.,  Hamana T.,  Kayo I.,   Futamase T.,  2002, \mn@doi [The
  Astrophysical Journal] {10.1086/340048}, 571, 638

\bibitem[\protect\citeauthoryear{{Tsaprazi}, {Nguyen}, {Jasche}, {Schmidt}  \&
  {Lavaux}}{{Tsaprazi} et~al.}{2022}]{fbi_intrinsic_alignment}
{Tsaprazi} E.,  {Nguyen} N.-M.,  {Jasche} J.,  {Schmidt} F.,   {Lavaux} G.,
  2022, \mn@doi [\jcap] {10.1088/1475-7516/2022/08/003}, \href
  {https://ui.adsabs.harvard.edu/abs/2022JCAP...08..003T} {2022, 003}

\bibitem[\protect\citeauthoryear{Vicinanza, Cardone, Maoli, Scaramella, Er  \&
  Tereno}{Vicinanza et~al.}{2019}]{Vicinanza_2019}
Vicinanza M.,  Cardone V.~F.,  Maoli R.,  Scaramella R.,  Er X.,   Tereno I.,
  2019, \mn@doi [Physical Review D] {10.1103/physrevd.99.043534}, 99

\bibitem[\protect\citeauthoryear{Wiener}{Wiener}{1949}]{wiener}
Wiener N.,  1949, Extrapolation, interpolation, and smoothing of Stationary
  time series: With engineering applications.
MIT Press

\bibitem[\protect\citeauthoryear{Xavier, Abdalla  \& Joachimi}{Xavier
  et~al.}{2016}]{flask}
Xavier H.~S.,  Abdalla F.~B.,   Joachimi B.,  2016, \mn@doi [Monthly Notices of
  the Royal Astronomical Society] {10.1093/mnras/stw874}, 459, 3693

\bibitem[\protect\citeauthoryear{Yiu, Fluri  \& Kacprzak}{Yiu
  et~al.}{2021}]{kids_gan}
Yiu T. W.~H.,  Fluri J.,   Kacprzak T.,  2021, A tomographic spherical mass map
  emulator of the KiDS-1000 survey using conditional generative adversarial
  networks, \mn@doi{10.48550/ARXIV.2112.12741}, \url
  {https://arxiv.org/abs/2112.12741}

\bibitem[\protect\citeauthoryear{{Z{\"u}rcher} et~al.,}{{Z{\"u}rcher}
  et~al.}{2022}]{desy3_peaks}
{Z{\"u}rcher} D.,  et~al., 2022, \mn@doi [\mnras] {10.1093/mnras/stac078},
  \href {https://ui.adsabs.harvard.edu/abs/2022MNRAS.511.2075Z} {511, 2075}

\makeatother
\end{thebibliography}

%\begin{thebibliography}{99}
%\end{thebibliography}

%%%%%%%%%%%%%%%%%%%%%%%%%%%%%%%%%%%%%%%%%%%%%%%%%%

%%%%%%%%%%%%%%%%% APPENDICES %%%%%%%%%%%%%%%%%%%%%

\appendix

\section{Chain Length Validation}
\label{sec:chain_length}

\begin{figure*}
    \centering
    \includegraphics[width=\textwidth]{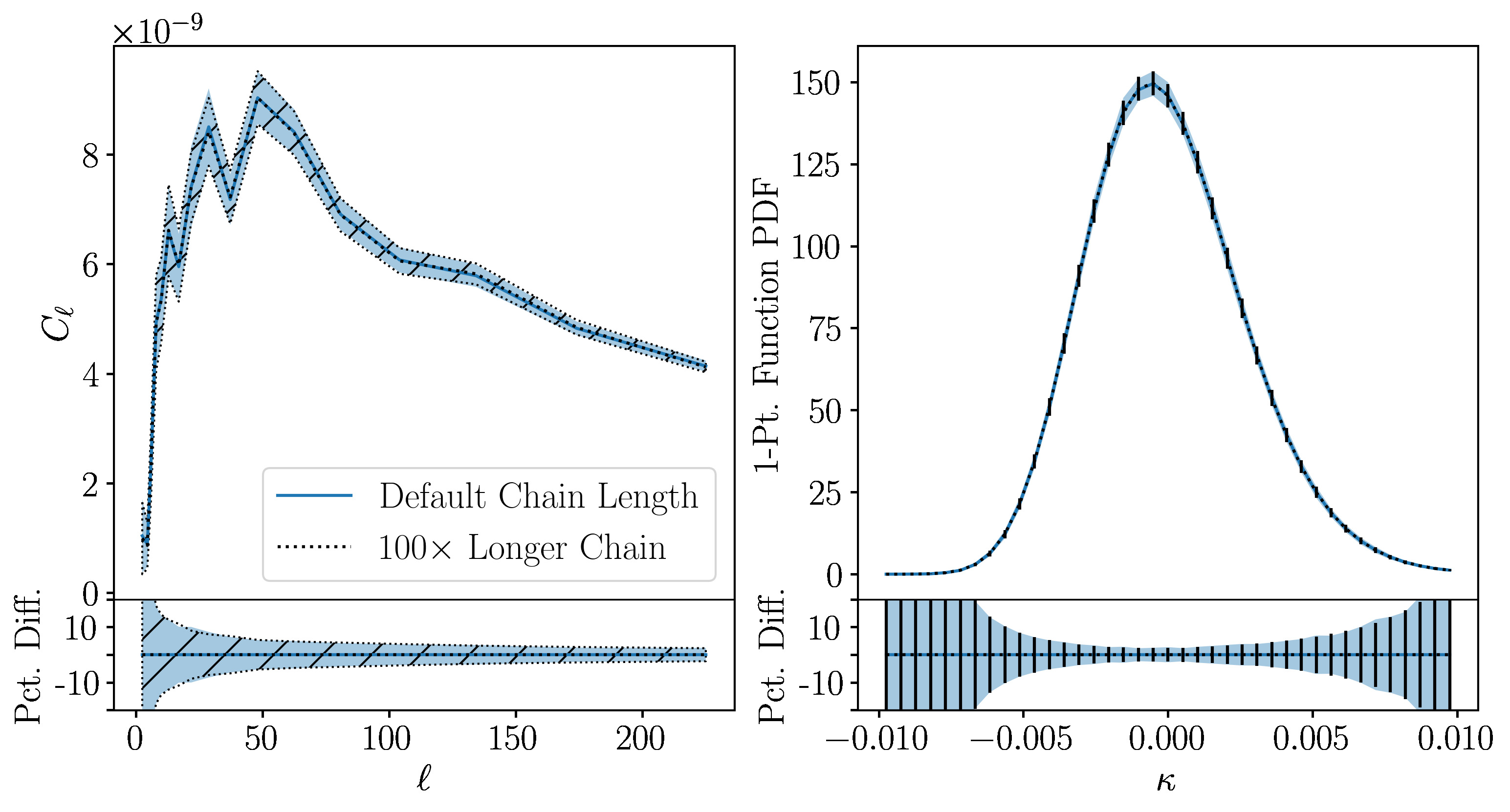}
    \caption[Chain length validation.]{Comparison between our default chain length of 200 samples and a longer chain with 20,000 samples over two-point (left) and one-point statistics (right) of \karmma{} samples for a single mock realization. The error bars represent the $1\sigma$ uncertainty from the posterior. It is clear that, despite the small number of samples, the statistical properties of the sample maps are well converged.}
    \label{fig:long_chain}
\end{figure*}

Throughout this work we use a default chain length of 200 samples for each mock realization. It is therefore unlikely that our maps fully sample the posterior space of mass maps. In fact, due to the extreme dimensionality of the problem (e.g., $\sim$ 2.5 million parameters for $\Nside=512$ maps), it is virtually impossible to sample the entire parameter space adequately. However, we do find that the statistical properties of the sampled mass maps are well converged. To verify this claim, for one of our mock realizations at $N_\mathrm{side}=128$ we run a chain where we generate 20,000 samples.  This chain is 100 times longer than our default chains. We then compare the recovered statistical properties from the default and extended chain lengths. Figure \ref{fig:long_chain} shows the one and two-point statistics of the sample maps for each chain length. From this plot, it is clear that our default chain length of 200 samples is adequate and accurately recovers the statistical properties from the extended runs. As stated previously, we attribute these results to a large survey footprint that effectively contains many independent sky patches.

%%%%%%%%%%%%%%%%%%%%%%%%%%%%%%%%%%%%%%%%%%%%%%%%%%

% Don't change these lines
\bsp	% typesetting comment
\label{lastpage}
\end{document}